\documentclass[12pt]{article} 
\usepackage{amsmath,amssymb}

\newcommand{\dps}{\displaystyle }
\newcommand{\e }{\varepsilon }

\newcommand{\al }{\alpha }
\newcommand{\de }{\delta }

\newcommand{\ket }{\rangle }
\newcommand{\bra }{\langle }
\newcommand{\ga }{\gamma }
\newcommand{\la }{\lambda }
\newcommand{\Hw }{H_{\hbox{w}}} 
\newcommand{\im}{\hbox{Im}}
\newcommand{\re}{\hbox{Re}}

\newcommand{\De }{\Delta }

\newcommand{\kob }{\overline{K^0}}
\newcommand{\ko }{K^0}

\begin{document}
\begin{flushright}
NUP-A-99-15\\September 1999
\end{flushright}
~\\ ~\\ 
\begin{center}
\Large{Direct $CP$, $T$ and/or $CPT$ violations in the $\ko-\kob$ system \\
- Implications of the recent KTeV results on $2\pi$ decays -}
\end{center} 
~\\ ~\\ 
\begin{center}
Yoshihiro Takeuchi\footnote{E-mail address: yytake@phys.cst.nihon-u.ac.jp} and S. Y. Tsai\footnote{E-mail address: tsai@phys.cst.nihon-u.ac.jp}
\end{center}

\begin{center}
{\it Atomic Energy Research Institute and Department of Physics \\ 
College of Science and Technology, Nihon University \\ 
Kanda-Surugadai, Chiyoda-ku, Tokyo 101-8308, Japan }
\end{center}
~\\ ~\\ ~\\ ~\\

\begin{abstract}
~~~~The recent results on the $CP$ violating parameters $\re(\e'/\e)$ and $\De\phi \equiv \phi_{00}-\phi_{+-}$ reported by the KTeV Collaboration are analyzed with a view to constrain $CP$, $T$ and $CPT$ violations in a decay process. Combining with some relevant data compiled by the Particle Data Group, we find $\re(\e_2-\e_0) = (0.85 \pm 3.11) \times 10^{-4}$ and $\im(\e_2-\e_0) = (3.2 \pm 0.7) \times 10^{-4}$, where $\re(\e_I)$ and $\im(\e_I)$ represent respectively $CP/CPT$ and $CP/T$ violations in decay of $\ko$ and $\kob$ into a $2\pi$ state with isospin $I$.  \\ \\ \\ \\
PACS 11.30.Er, 13.20.Eb, 13.25.Es

\end{abstract}

\newpage
Although it has been well established since 1964 [1] that $CP$ symmetry is violated in the $\ko-\kob$ system, origin or mechanism of $CP$ violation is not well understood yet on the one hand and no evidence of $CP$ violation has been established in any other systems or processes on the other hand. Experimental, phenomenological and theoretical studies of this and related (i.e., $T$ and $CPT$) symmetries need to be continued with much efforts.

The KTeV Collaboration [2] recently reported
%
%
\begin{subequations}
\begin{eqnarray}
\re(\e'/\e) &=& (2.80 \pm 0.41) \times 10^{-3}~, \\
\De\phi &=& (0.09 \pm 0.46)^{\circ}~,
\end{eqnarray}
\end{subequations}
and claimed that the fact $\re(\e'/\e) \neq 0$ definitively established the existence of $CP$ violation in a decay process. In the present note, we like to analyse in detail what the KTeV results imply and to see in particular how well $CPT$ symmetry is tested compared to $T$ symmetry. \\
\\
{\it The $\ko-\kob$ mixing and $2\pi$ decays} \\

Let $|\ko\ket$ and $|\kob\ket$ be eigenstates of the strong interaction with strangeness $S=+1$ and $-1$, related to each other by $(CP)$ and $(CPT)$ operations as [3,4]
%
%
\begin{equation}
(CP)|\ko\ket = e^{i\al_K}|\kob\ket~,\qquad (CPT)|\ko\ket = e^{i\beta_K}|\kob\ket ~,
\end{equation}
where $\al_K$ and $\beta_K$ are arbitrary real parameters. When the weak interaction $\Hw$ is switched on, $\ko$ and $\kob$ decay into other states, generically denoted as $n$, and get mixed. The states with definite mass ($m_{S,L}$) and width ($\ga_{S,L}$; $\ga_S > \ga_L$ by definition) are linear combinations of $\ko$ and $\kob$:
%
%
\begin{subequations}
\begin{eqnarray}
|K_S\ket &=&\frac{1}{\sqrt{|p_S|^2+|q_S|^2}}(p_S|\ko \ket + q_S|\kob \ket) ~, \\|K_L\ket &=&\frac{1}{\sqrt{|p_L|^2+|q_L|^2}}(p_L|\ko \ket - q_L|\kob \ket) ~. 
\end{eqnarray}
\end{subequations}
The ratios of the mixing parameters, $q_{S,L}/p_{S,L}$, as well as $\la_{S,L} \equiv m_{S,L}-i\ga_{S,L}/2$, are related to $\Hw$; the explicit expressions can be found in the literature [3,5]. We are interested in $2\pi$ decays and specifically in the following quantities:
%
%
\begin{subequations}
\begin{eqnarray}
\eta _{+-}=|\eta _{+-}|e^{i\phi _{+-}} &\equiv& \frac{\bra \pi ^+\pi ^-,\mbox{outgoing}|\Hw |K_L\ket }{\bra \pi ^+\pi ^-,\mbox{outgoing}|\Hw |K_S\ket }~, \\
\eta _{00}=|\eta _{00}|e^{i\phi _{00}} &\equiv& \frac{\bra \pi ^0\pi ^0,\mbox{outgoing}|\Hw |K_L\ket }{\bra \pi ^0\pi ^0,\mbox{outgoing}|\Hw |K_S\ket }~,
\end{eqnarray}
\end{subequations}
%
%
\begin{equation}
r \equiv \frac{\ga_S(\pi^+\pi^-)-2\ga_S(\pi^0\pi^0)}{\ga_S(\pi^+\pi^-)+\ga_S(\pi^0\pi^0)}, 
\end{equation}
where $\ga_{S,L}(n)$ denotes the partial width for $K_{S,L}$ to decay into the final state $n$. \\
\\
{\it Parametrization and conditions imposed by $CP$, $T$ and $CPT$ symmetries} \\

We shall parametrize $q_S/p_S$ and $q_L/p_L$ as [3]
%
%
\begin{subequations}
\begin{eqnarray}
	\frac{\dps{q_S}}{\dps{p_S}}=e^{i\al_K}\frac{\dps{1-\e-\de}}{\dps{1+\e+\de}}~, \\
	\frac{\dps{q_L}}{\dps{p_L}}=e^{i\al _K}\frac{\dps{1-\e+\de}}{\dps{1+\e-\de}}~,\end{eqnarray}
\end{subequations}
and the amplitudes for $\ko$ and $\kob$ to decay into 2$\pi$ states with isospin $I$ = 0 or 2 as [3,6]
%
%
\begin{subequations}
\begin{eqnarray}
\bra(2\pi)_I|\Hw|\ko\ket &=& F_I(1+\e_I)e^{i\al_K/2}~, \\
\bra(2\pi)_I|\Hw|\kob\ket &=& F_I(1-\e_I)e^{-i\al_K/2}~.
\end{eqnarray}
\end{subequations}

Our parametrization is very unique in that it is invariant under rephasing of the initial states, $|\ko\ket$ and $|\kob\ket$. It is however not invariant under rephasing of the final states, $|(2\pi)_I\ket$. By making use of the phase ambiguity, one may, without loss of generality, set [6]
%
%
\begin{equation}
\im(F_I) = 0~.
\end{equation}
One readily verify [3,6] that $CP$, $T$ and $CPT$ symmetries impose such conditions as
%
%
\begin{equation}
\left. 
	\begin{array}{ccl}
	CP~\mbox{symmetry} &:& \e = 0,~\de = 0,~\e_I = 0~; \\ 
	T~\mbox{symmetry} &:& \e = 0,~\im(\e_I) = 0~; \\ 
	CPT~\mbox{symmetry} &:& \de = 0,~\re(\e_I) = 0~. 
	\end{array}
\right. 
\end{equation}
Observed and expected smallness of symmetry violation allows one to treat all these parameters as small. \\
\\
{\it Formulae relevant for analysis} \\

Defining
%
%
\begin{subequations}
\begin{eqnarray}
\eta_I = |\eta_I|e^{i\phi_I} \equiv \frac{\bra(2\pi)_I|\Hw|K_L\ket}{\bra(2\pi)_I|\Hw|K_S\ket}~, \\
\omega \equiv \frac{\bra(2\pi)_2|\Hw|K_S\ket}{\bra(2\pi)_0|\Hw|K_S\ket}~,
\end{eqnarray}
\end{subequations}
one finds [7,8], from Eqs.(3a,b), (6a,b) and (7a,b),
%
%
\begin{subequations}
\begin{eqnarray}
\eta_I &=& \e - \de + \e_I~, \\
\omega &=& \re(F_2)/\re(F_0)~,
\end{eqnarray}
\end{subequations}
and, by means of isospin decomposition,
%
%
\begin{subequations}
\begin{eqnarray}
\eta_{+-} = \eta_0 + \e'~, \\
\eta_{00} = \eta_0 - 2\e'~,
\end{eqnarray}
\end{subequations}
%
%
\begin{equation}
r = 4\re(\omega')~,
\end{equation}
where
%
%
\begin{subequations}
\begin{eqnarray}
\e' &\equiv& (\eta_2-\eta_0)\omega'~, \\
\omega' &\equiv& \frac{1}{\sqrt{2}}\omega e^{i(\de_2-\de_0)}~,
\end{eqnarray}
\end{subequations}
$\de_I$ being the S-wave $\pi \pi$ scattering phase shift for the isospin $I$ state at an energy of the rest mass of $\ko$. Note that we have treated $\omega'$, which is a measure of deviation from the $\De I=1/2$ rule, as well as a small quantity. From Eqs.(12a,b), it follows that
%
%
\begin{equation}
\eta _{00}/\eta _{+-} = 1 - 3\e'/\eta_0~,
\end{equation}
or
%
%
\begin{subequations}
\begin{eqnarray}
\re(\e'/\eta_0) &=& (1/3)(1 - |\eta_{00}/\eta_{+-}|)~, \\
\im(\e'/\eta_0) &=& -(1/3)\De\phi~,
\end{eqnarray}
\end{subequations}
where
%
\begin{equation}
\De\phi \equiv \phi_{00} - \phi_{+-}~.
\end{equation}
\\
{\it Implications of the KTeV results} \\

With the help of the formulae derived above, we now look into implications of the latest results reported by the KTeV Collaboration [2]. We first note that, since $\e$ in their notation corresponds exactly to $\eta_0$ in our notation,\footnote{For the correspondence between our parametrization and the (more conventional) rephasing-dependent parametrizations, see [3,8].} their results (1a,b) give, either immediately or with the help of Eqs.(16a,b),
%
%
\begin{subequations}
\begin{eqnarray}
\re(\e'/\eta_0) &=& (~~2.80 \pm 0.41) \times 10^{-3}~, \\
\im(\e'/\eta_0) &=& (-0.52 \pm 2.68) \times 10^{-3}~, \\
|\eta_{00}/\eta_{+-}| &=& 0.9916 \pm 0.0012~.
\end{eqnarray}
\end{subequations}
From Eqs.(11a) and (14a), we immediately conclude that $\e' \neq 0$ implies that either $\e_0$ or $\e_2$ (or both) is $\neq 0$,\footnote{Note that the reverse is however not necessarily true; a nonvanishing but equal value for both $\e_0$ and $\e_2$ could yield $\e'=0$.} confirming the assertion that the KTeV result on $\re(\e'/\e)$ established the existence of $CP$ violation in a decay process [2].

To go one step further, we need to know the value of $\eta_0$. Since the KTeV collaboration has not yet reported their results on $\eta_{+-}$ and $\eta_{00}$ separately, we shall input the PDG [9] values for $\eta_{+-}$,
%
%
\begin{subequations}
\begin{eqnarray}
|\eta_{+-}| &=& (2.285 \pm 0.019) \times 10^{-3}~, \\
\phi_{+-} &=& (43.5 \pm 0.6)^{\circ}~,
\end{eqnarray}
\end{subequations}
along with Eqs.(1b) and (18c), into
%
%
\begin{equation}
\eta_0 \simeq (2/3)\eta_{+-} + (1/3)\eta_{00},
\end{equation}
which follows from Eqs.(12a,b), to get
%
%
\begin{subequations}
\begin{eqnarray}
|\eta_0| &=& (2.28 \pm 0.02) \times 10^{-3}~, \\
\phi_0 &=& (43.53 \pm 0.94)^{\circ}~.
\end{eqnarray}
\end{subequations}
We shall also use the PDG [9] values for $\ga_S(\pi^+\pi^-)$ and $\ga_S(\pi^0\pi^0)$  to get, with the help of Eqs.(5) and (13),
%
\begin{equation}
\re(\omega') = (1.46 \pm 0.16) \times 10^{-2}~.
\end{equation}
In order to interpret Eqs.(18a,b), we derive from Eqs.(14a,b), with the aid of Eqs.(11a,b),
%
%
\begin{equation}
\e'/\eta_0 = -i\re(\omega')(\e_2-\e_0)e^{-i\De\phi'}/[|\eta_0|\cos(\de_2-\de_0)]~,
\end{equation}
or
%
%
\begin{equation}
\e_2-\e_0 = i(\e'/\eta_0)|\eta_0|\cos(\de_2-\de_0)e^{i\De\phi'} /\re(\omega')~,
\end{equation}
where
%
%
\begin{equation}
\De\phi' \equiv \phi_0 - \de_2 + \de_0 - \pi/2~.
\end{equation}
Inputting Eqs.(18a,b), (21a,b) and (22), and $\de_2-\de_0$ as well, into Eq.(24), we are able to derive constraints to $\re(\e_2-\e_0)$ and $\im(\e_2-\e_0)$:%
%
%
\begin{subequations}
\begin{eqnarray}
\re(\e_2-\e_0) = (0.85 \pm 3.11) \times 10^{-4}~, \\
\im(\e_2-\e_0) = (3.2~~ \pm 0.7~~) \times 10^{-4}~,
\end{eqnarray}
\end{subequations}
where, as $\de_2-\de_0$, we have tentatively used the Chell-Olsson value, $(-42 \pm 4)^{\circ}$ [10]. \\
\\
{\it Discussion} \\

Our result (26b) indicates that a combination of the parameters which signal direct $CP$ and $T$ violations, $\im(\e_2-\e_0)$, is definitely nonzero and of the order of $10^{-4}$. The other result (26a) on the other hand indicates that a combination of the parameters which signal direct $CP$ and $CPT$ violations, $\re(\e_2-\e_0)$, is not well determined yet; though consistent with being zero, a value comparable to or even larger than $\im(\e_2-\e_0)$ is not ruled out.

If, instead of the KTeV values, Eqs.(1a,b), one inputs the PDG [9] values also for  $|\eta_{00}/\eta_{+-}|$ and $\De\phi$,
%
%
\begin{subequations}
\begin{eqnarray}
|\eta_{00}/\eta_{+-}| &=& 0.9956 \pm 0.0023~, \\
\De\phi &=& (-0.1 \pm 0.8)^{\circ}~,
\end{eqnarray}
\end{subequations}
one will get
%
%
\begin{subequations}
\begin{eqnarray}
\re(\e'/\eta_0) &=& (1.5 \pm 0.8) \times 10^{-3}~, \\
\im(\e'/\eta_0) &=& (0.6 \pm 4.7) \times 10^{-3}~,
\end{eqnarray}
\end{subequations}
and
%
%
\begin{subequations}
\begin{eqnarray}
\re(\e_2-\e_0) &=& (-0.56 \pm 5.45) \times 10^{-4}~, \\
\im(\e_2-\e_0) &=& (~~1.8~~ \pm 1.0~~) \times 10^{-4}~.
\end{eqnarray}
\end{subequations}

With the help of the Bell-Steinberger relation [11], one may derive constraints to the "indirect" and "mixed" $CP$, $T$ and/or $CPT$ violating parameters [7,8,12,13]. It turns out that the values of the direct $CP/T$ violating parameter we have obtained, Eqs.(26b) and(29b), are almost one order smaller than those of the indirect and mixed $CP/T$ violating parameters, $\re(\e)$ and $\im(\e+\e_0)$, while the constraints on the direct $CP/CPT$ violating parameter we have found, Eqs.(26a) and (29a), are roughly one order weaker than those on the indirect and mixed $CP/CPT$ violating parameters, $\im(\de)$ and $\re(\de-\e_0)$.\footnote{$\e_0$ and $\e_2$ ($\e$ and $\de$) are referred to as a direct (indirect) parameter here. Note that, as emphasized in [3], classification of symmetry-violating parameters into "direct" and "indirect" ones makes sense only when they are defined in a rephasing-invariant way, i.e., in such a way that they are invariant under rephasing of $|\ko\ket$ and $|\kob\ket$.}

To conclude, we recall that the numerical results (26a,b) and (29a,b) depend much on the value of $\de_2-\de_0$, and that this quantity, which features strong interaction effects, is still not well determined. In order to obtain a better constraint on $\e_2-\e_0$, a better determination of $\de_2-\de_0$, along with a more precise measurement of $\De\phi$, are required.

\section*{Acknowledgements}

~~~~We are grateful to Professor T.Yamanaka for a discussion on the results and details of the KTeV experiment.

\newpage


\begin{thebibliography}{100}

\bibitem{1} J.H.Christenson et al., Phys.Rev.Lett. 13, 138 (1964). \\
     T.T.Wu and C.N.Yang, Phys.Rev.Lett. 13, 380 (1964). 
\bibitem{2} A.Alavi-Harati et al., Phys.Rev.Lett. 83, 22 (1999). \\
     E.Blucher, talk presented at Rencontres de Moriond (March 15, 1999).
\bibitem{3} K.Kojima, W.Sugiyama and S.Y.Tsai, Prog.Theor.Phys. 95, 913 (1996). \\
     S.Y.Tsai, Mod.Phys.Lett. A 11, 2941 (1996).
\bibitem{4} W.J.Mantke, preprint (MPI-PhT/94-98).
\bibitem{5} T.D.Lee, R.Oehme and C.N.Yang, Phys.Rev. 106, 340 (1957). \\
     T.D.Lee and C.S.Wu, Ann.Rev.Nucl.Sci. 16, 511 (1966).
\bibitem{6} S.Y.Tsai, in Proceedings of the 8'th B-Physics International Workshop (Kawatabi, Miyagi, October 29-31,1998), edited by K.Abe et al., p.95. \\
     Y.Kouchi, Y.Takeuchi and S.Y.Tsai, preprint (NUP-A-99-14, hep-ph/\\9908201).
\bibitem{7} K.Kojima, A.Shinbori, W.Sugiyama and S.Y.Tsai, Prog.Theor.Phys. 97, 103 (1997). \\
     A.Shinbori, N.Hashimoto, K.Kojima, T.Mochizuki, W.Sugiyama and S.Y.Tsai, in Proceedings of the 5'th KEK Meeting on CP Violation and its Origin (KEK, March 6-7), edited by K.Hagiwara (KEK Proceedings 97-12, October 1997), p.181.
\bibitem{8} Y.Kouchi, master's thesis (in Japanese). \\
     Y.Kouchi, A.Shinbori, Y.Takeuchi and S.Y.Tsai, in Proceedings of the International Workshop on Fermion Masses and CP Violation (Hiroshima, March 5-6,1998), edited by T. Morozumi and T.Muta (Hiroshima University, 1998), p.79.
\bibitem{9} Particle Data Group, Eur.Phys.J. C 3, 1 (1998).
\bibitem{10} E.Chell and M.G.Olsson, Phys.Rev. D 48, 4076 (1993).
\bibitem{11} J.S.Bell and J.Steinberger, in Proceedings of the Oxford International Conference on Elementary Particles, edited by R.G.Moorhouse et al.(Lutherford Laboratory, 1965), p.195.
\bibitem{12} Y.Kouchi, Y.Takeuchi and S.Y.Tsai, in preparation.
\bibitem{13} CPLEAR Collaboration, A.Apostolakis et al., Phys.Lett. B 456, 297 (1999).
\end{thebibliography}
\end{document}